\documentclass{article}
\usepackage{amsfonts}
\usepackage{amssymb}
\usepackage{epsfig}

\def\preline{\\[-8pt]}                 
\def\postline{\\[-6pt]}                
\def\firsthline{\hline\postline}
\def\midhline{\preline\hline\postline}
\def\lasthline{\preline\hline}

\title{Effective masses of diquarks}
\author{P. Maris\thanks{\textit{E-mail address:} 
pmaris@unity.ncsu.edu}\\
\footnotesize{Dept. of Physics, 
North Carolina State University, 
Raleigh,  NC 27695-8202}}

\begin{document}

\maketitle
\begin{abstract}
We study meson and diquark bound states using the rainbow-ladder
truncation of QCD's Dyson--Schwinger equations.  The infrared strength
of the rainbow-ladder kernel is described by two parameters.  The
ultraviolet behavior is fixed by the one-loop renormalization group
behavior of QCD, which ensures the correct asymptotic behavior of the
Bethe--Salpeter amplitudes and brings important qualitative benefits.
The diquark with the lowest mass is the scalar, followed by the
axialvector and pseudoscalar diquark.  This ordering can be
anticipated from the meson sector.
\end{abstract}

\section{\label{sec:intro}
Introduction}
Mesons are color-singlet bound states of a quark and an antiquark.  
In addition to quark-antiquark bound states, one could also ask the
question whether or not there are quark-quark bound states, by
studying the corresponding Bethe--Salpeter equation [BSE] for bound
states.  Such states are of course not colorless in QCD, and are
therefore expected to be confined, if they exist at all.
Nevertheless, the masses of these states can serve as an indication
for the relevant mass scales of quark-quark correlations.  Such
diquark correlations could play a role inside baryons: two quarks
bound in a color antitriplet configuration can couple with a quark to
form a color-singlet baryon.  Indeed, recent studies of baryons as
bound states of a quark and a (confined) diquark are quite successful
in describing baryons~\cite{rtcahill,bentz,tuebingen,argonne}.  Also
some lattice simulations~\cite{Hess:1998sd} indicate the existence of
correlations in diquark channels.

Here we report ground state meson and color-antitriplet diquark masses
based on the rainbow-ladder truncation of the set of Dyson--Schwinger
equations [DSEs]~\cite{Roberts:2000aa,Alkofer:2000wg}.  This covariant
approach accommodates quark confinement and implements QCD's one-loop
renormalization group behavior.  The model we apply has previously
been shown to give an efficient description of the masses and
electroweak decay constants of the light pseudoscalar and vector
mesons~\cite{Maris:1997tm,Maris:1999nt} and their
interactions~\cite{Maris:emf,Ji:2001pj,Maris:2001am,Maris:2002mz}.
This calculation will provide information to guide the improvement of
quark-diquark models of baryons and assists in developing an intuitive
understanding of lattice simulations.

\section{\label{sec:DSEapproach}
Dyson--Schwinger Equations}
The DSE for the renormalized quark propagator in Euclidean space is
\begin{eqnarray}
  S(p)^{-1} &=& i \, Z_2\, /\!\!\!p + Z_4\,m_q(\mu) 
	+  Z_1  \int_q^\Lambda\!  g^2 D_{\mu\nu}(k) \, 
        {\textstyle\frac{\lambda^i}{2}} \gamma_\mu \, 
	S(q) \, \Gamma^i_\nu(q,p) \; ,
\label{gendse}
\end{eqnarray}
where $D_{\mu\nu}(k=p-q)$ is the dressed-gluon propagator,
$\Gamma^i_\nu(q,p)$ the dressed-quark-gluon vertex with color-octet
index $i=1,\ldots,8$, and $Z_2$ and $Z_4$ are the
quark wave-function and mass renormalization constants.  The notation
\mbox{$\int^\Lambda_q \equiv \int^\Lambda d^4 q/(2\pi)^4$} stands for
a translationally invariant regularization of the integral, with
$\Lambda$ being the regularization mass-scale.  The regularization is
removed at the end of all calculations, by taking the limit
\mbox{$\Lambda \to \infty$}.  We use the Euclidean metric where
\mbox{$\{\gamma_\mu,\gamma_\nu\} = 2\delta_{\mu\nu}$},
\mbox{$\gamma_\mu^\dagger = \gamma_\mu$} and \mbox{$a\cdot b =
\sum_{i=1}^4 a_i b_i$}.  The most general solution of
Eq.~(\ref{gendse}) has the form \mbox{$S(p)^{-1} = i /\!\!\! p A(p^2)
+$} \mbox{$B(p^2)$} and is renormalized at spacelike $\mu^2$ according
to \mbox{$A(\mu^2)=1$} and \mbox{$B(\mu^2)=m_q(\mu)$} with $m_q(\mu)$
being the current quark mass.

\subsection{\label{sec:mesons}
Meson Bethe--Salpeter Equation}
\begin{figure}[hbt]
\begin{center}
\epsfig{file=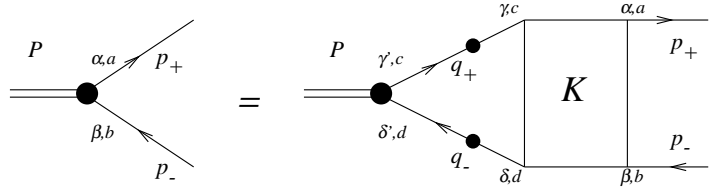,width=100mm}
\end{center}
\caption[]{Meson Bethe--Salpeter Equation.\label{fig:bseM}}
\end{figure}

Mesons are color-singlet quark-antiquark bound states.  They are
solutions of the homogeneous Bethe--Salpeter equation [BSE] for $q
\bar{q}$ bound states
\begin{eqnarray}
 \Gamma^{\alpha\beta}_M(p_+,p_-) \delta_{ab} &=& \int_q^\Lambda\! 
	K_{ac;db}^{\alpha\gamma;\delta\beta}(p_+,q_+;q_-,p_-) 
\nonumber \\ && 
	S^{\gamma\gamma'}(q_+) \, 
	\Gamma^{\gamma'\delta'}_M(q_+,q_-) \delta_{cd}
	\, S^{\delta'\delta}(q_-) \, ,
\label{eq:MesonBSE}
\end{eqnarray}
where $p_+ = p + \eta P$ and $p_- = p - (1-\eta) P$ are the outgoing
and incoming quark momenta respectively, and $q_\pm$ is defined
similarly, see also Fig.~\ref{fig:bseM}.  
The greek superscripts are spinor indices, and the roman subscripts
are color indices running from 1 to 3; for simplicity we consider one
flavor here.  The kernel $K$ is the renormalized, amputated $q\bar q$
scattering kernel that is irreducible with respect to a pair of $q\bar
q$ lines.

This equation has solutions at discrete values of $P^2 = -m_M^2$,
where $m_M$ is the meson mass.  Together with the canonical
normalization condition for $q\bar q$ bound states, it completely
determines $\Gamma_M$, the bound state Bethe--Salpeter amplitude
[BSA].  The different types of mesons, such as pseudo-scalar, vector,
etc. are characterized by different Dirac structures.  The most
general decomposition for pseudoscalar bound states
is
\begin{eqnarray}
\label{genpion}
\Gamma_{PS}(q_+,q_-) &=& \gamma_5 \big[ i E(q^2,q\cdot P;\eta) 
        + \;/\!\!\!\! P \, F(q^2,q\cdot P;\eta) 
\nonumber \\ && {}
        + \,/\!\!\!k \, G(q^2,q\cdot P;\eta) 
        + \sigma_{\mu\nu}\,P_\mu q_\nu \,H(q^2,q\cdot P;\eta) \big]\,,
\end{eqnarray}
where the invariant amplitudes $E$, $F$, $G$ and $H$ are Lorentz
scalar functions of $q^2$ and $q\cdot P$.  Note also that these
amplitudes explicitly depend on the momentum partitioning parameter
$\eta$.  However, so long as Poincar\'e invariance is respected, the
resulting physical observables are independent of this
parameter~\cite{Maris:1997tm,Maris:emf}.  If $\eta = \frac{1}{2}$,
these amplitudes are appropriately odd or even in the charge parity
odd quantity $q\cdot P$ for charge eigenstates.  In the case of the
$0^{-+}$ pion, for example, the amplitude $G$ is odd in $q\cdot P$,
the others are even.

The most general decomposition for scalar mesons can be obtained from
that for a pseudoscalar mesons by dropping the $\gamma_5$.  For the
$0^{++}$, the amplitude $F$ is odd in $q\cdot P$, the others are even,
if we use the same notation as for the pseudoscalars.  A vector meson
has more Dirac structures: a massive vector meson being transverse,
the general decomposition of a such a BSA requires eight
covariants~\cite{Maris:1999nt}, the dominant structure being
\begin{eqnarray}
\Gamma^{\hbox{\scriptsize{dom}}}_\mu(q_+,q_-) &=& 
	\left(\delta_{\mu\nu} - \frac{P_\mu P_\nu}{P^2}\right) 
	\, \gamma_\nu  \; V_1(q^2,q\cdot P;\eta)  \;.
\label{vecBSAdom}
\end{eqnarray}
For the $1^{--}$ $\rho$ meson, the function $V_1(q^2,q\cdot
P;\eta=\frac{1}{2})$ is even in $q\cdot P$.

\subsection{\label{sec:diquarks}
Diquark bound states}
Diquarks are quark-quark correlations.  In QCD, with $N_c=3$, these
states are necessarily colored and therefore believed to be confined.
Two quarks can be coupled in either a color sextet or a color
antitriplet.  Single gluon exchange leads to an (effective)
interaction that is attractive for diquarks in a color antitriplet
configuration, but the interaction is repulsive in the color sextet
channel~\cite{Cahill:1987qr,Hecht:2000jh}.  Furthermore, it is the
diquark in a color antitriplet that can couple with a quark to form a
color-singlet baryon.  Thus we will only consider in diquarks in a
color antitriplet configuration.

\begin{figure}[hbt]
\begin{center}
\epsfig{file=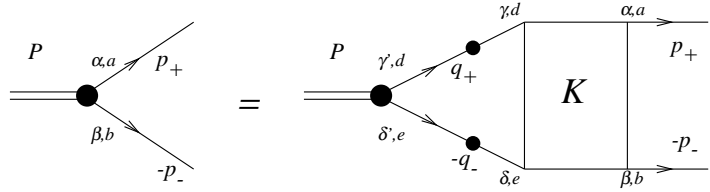,width=100mm}
\end{center}
\caption[]{Diquark Bethe--Salpeter equation.\label{fig:bseD}}
\end{figure}

Using the antisymmetric tensor $\epsilon_{abc}$ as a representation of
the antitriplet~\cite{Alkofer:mv}, the corresponding diquark bound
states can be described by solutions of the homogeneous BSE
\begin{eqnarray}
 \Gamma^{\alpha\beta}_D(p_+,-p_-) \epsilon_{abc} &=& \int_q^\Lambda\! 
	K_{ad;be}^{\alpha\gamma;\beta\delta}(p_+,q_+;-p_-,-q_-) 
\nonumber \\ && 
        S^{\gamma\gamma'}(q_+) \, 
	\Gamma^{\gamma'\delta'}_M(q_+,-q_-) \, \epsilon_{dec}
	\, S^{\delta\delta'}(-q_-) \, ,
\end{eqnarray}
where $p_+ = p + \eta P$ and $-p_- = -p + (1-\eta) P$ are now {\em
both outgoing} quark momenta (and similarly for $q_+$ and $-q_-$).
This is schematically depicted in Fig.~\ref{fig:bseD}; note the
difference in the argument and the order of the indices of the second
quark propagator and of the kernel $K$ compared to the meson BSE,
Eq.~(\ref{eq:MesonBSE}).  
With the help of the charge conjugation matrix $C$, which satisfies
$C^2 = -1$ and $ C\,\gamma_\mu\,C \,=\, (\gamma_\mu)^T$, we can
rewrite the BSE for diquarks as~\cite{Cahill:1987qr}
\begin{eqnarray}
 \Big(\Gamma_D(p_+,-p_-)C\Big)^{\alpha\beta'} \epsilon_{abc} &=& 
	-\epsilon_{dec} \int_q^\Lambda\! 
	K_{ad;be}^{\alpha\gamma;\beta\delta}(p_+,q_+;-p_-,-q_-) 
	\, C^{\delta''\delta} \, C^{\beta\beta'}
\nonumber \\ && 
 	S^{\gamma\gamma'}(q_+) \, 
	\Big(\Gamma_D(q_+,-q_-)C\Big)^{\gamma'\delta'} \, 
	S(q_-)^{\delta'\delta''} \, .
\label{eq:DiqBSE}
\end{eqnarray}
This equation resembles the meson BSE; in fact, using a skeleton
expansion of the kernel $K$ one can show that $\Gamma_D(q_+,-q_-)C$
satisfies a BSE whose Dirac structure is identical to that of the
meson BSE for $\Gamma_M(q_+,q_-)$.  Only its color structure is
different, and that has important consequences~\cite{Bender:2002as}.

As in the case of mesons, the different types of diquarks are
characterized by different Dirac structures.  Since the intrinsic
parity of a quark-quark pair is opposite to that of a quark-antiquark
pair, a scalar diquark BSA, or rather
$\Gamma_{D,{\hbox{\scriptsize{scal}}}}(q_+,-q_-)C$, has exactly the
same form as a pseudoscalar meson BSA, Eq.~(\ref{genpion}).
Similarly, an axialvector diquark BSA has the same decomposition as a
vector meson BSA, and a pseudoscalar diquark BSA has the same form as
a scalar meson BSA.

\subsection{\label{sec:ladder}
Two-body bound states in rainbow-ladder truncation}
For practical calculations, we utilize the rainbow truncation of the
quark DSE
\begin{eqnarray}
Z_1 \, g^2 D_{\mu \nu}(k)\, \Gamma^i_\nu(q,p) &\rightarrow&
 {\cal G}(k^2)\, D_{\mu\nu}^{\rm free}(k)\, \gamma_\nu
                                        \textstyle\frac{\lambda^i}{2} \,,
\label{eq:lrainbowDSE}
\end{eqnarray}
in combination with the ladder truncation of the BSE
\begin{eqnarray}
\lefteqn{
  K_{ac;db}^{\alpha\gamma;\delta\beta}(p_+,q_+;q_-,p_-) \; \to}
\nonumber \\ &&
	-{\cal G}(k^2) \; 
        \left(\frac{\lambda^i}{2}\right)_{ac}
	\left(\gamma_\mu\right)^{\alpha\gamma} \; D_{\mu\nu}^{\rm free}(k) \;
        \left(\frac{\lambda^i}{2}\right)_{db}
	\left(\gamma_\nu\right)^{\delta\beta} \; ,
\label{eq:ladderK}
\end{eqnarray}
where $D_{\mu\nu}^{\rm free}(k=p-q=p_\pm-q_\pm)$ is the free gluon
propagator in Landau gauge.  These two truncations are consistent in
the sense that the combination produces vector and axialvector
vertices satisfying the respective Ward--Takahashi
identities~\cite{Maris:1997tm}.  In the axial case, this ensures that
in the chiral limit the ground state pseudoscalar mesons are the
massless Goldstone bosons associated with chiral symmetry breaking;
with nonzero current quark masses it leads to a generalization of the
Gell-Man--Oaks--Renner relation~\cite{Maris:1998hd}.  In the vector
case, this ensures conservation of the electromagnetic
current~\cite{Maris:emf}.

If we insert this ladder kernel, Eq.~(\ref{eq:ladderK}), in the meson BSE,
we get the ladder BSE for the meson BSA
\begin{eqnarray}
 \Gamma^{\alpha\beta}_M(p_+,p_-) &=& 
	-\frac{1}{3}\,
	{\rm Tr}_c\left[\frac{\lambda^i}{2} \, \frac{\lambda^i}{2} \right]
	\int_q^\Lambda\! {\cal G}(k^2) \; D_{\mu\nu}^{\rm free}(k)
\nonumber \\ &&
	\Big(\gamma_\mu \, S(q_+) \, \Gamma_M(q_+,q_-) \, 
	S(q_-) \, \gamma_\mu \Big)^{\alpha\beta} \; ,
\label{eq:ladderBSEm}
\end{eqnarray}
For the diquark bound states we obtain
\begin{eqnarray}
 \Big(\Gamma_D(p_+,-p_-) \, C\Big)^{\alpha\beta} \epsilon_{abc} &=& 
	\epsilon_{dec} 
        \left(\frac{\lambda^i}{2}\right)_{ad}
        \left(\frac{\lambda^i}{2}\right)_{be} \;
        \int_q^\Lambda\! {\cal G}(k^2) \; D_{\mu\nu}^{\rm free}(k)
\nonumber \\ &&
	\Big(\gamma_\mu \, S(q_+)  \, \Gamma_D(q_+,-q_-) \, 
	C \, S(q_-) \, \gamma_\mu \Big)^{\alpha\beta} \; .
\label{eq:ladderBSEd}
\end{eqnarray}
Comparing Eq.~(\ref{eq:ladderBSEm}) with Eq.~(\ref{eq:ladderBSEd}), we
see that in ladder truncation the only difference between the meson
BSE for $\Gamma_M$ and BSE for a color-antitriplet diquark, or rather,
for $\Gamma_D\,C$, lies in the color factors.  For the mesons, we have
${\rm Tr}_c [ \frac{\lambda^i}{2} \,\frac{\lambda^i}{2}] = 4$, and thus we
get a factor of $-4/3$ in front of the BSE integral.  For the diquarks
in a color antitriplet configuration on the other hand, we have
\begin{eqnarray}
 \left(\frac{\lambda^i}{2}\right)_{ad}
 \left(\frac{\lambda^i}{2}\right)_{be} \;
	\epsilon_{dec}
	&=&  -\frac{2}{3}\,\epsilon_{abc}  \;.
\label{eq:diqcol}
\end{eqnarray}
Thus, in {\em ladder truncation} the effective interaction for the
diquarks is reduced by a factor of two compared to the interaction in
the meson channel.  Beyond ladder truncation a more complex algebraic
relation persists between the meson and diquark
BSEs~\cite{Bender:2002as}.  We therefore expect that the diquarks with
the lowest mass are the scalar diquarks (the diquark partners of the
pseudoscalar mesons), followed by the axialvector and pseudoscalar
diquarks~\cite{Cahill:1987qr}.

The rainbow-ladder truncation is particularly suitable for the flavor
octet pseudoscalar mesons and for the vector mesons, since the
next-to-leading-order contributions in a quark-gluon skeleton graph
expansion have a significant amount of cancellation between repulsive
and attractive corrections~\cite{Bender:2002as,Bender:1996bb}.
However, the BSE for diquarks, and also the scalar meson BSE, receive
large repulsive corrections from these next-to-leading-order
contributions.  This will significantly change the scalar meson mass
and corresponding BSA; the diquark bound states found in ladder
truncation will in fact disappear altogether from the
spectrum~\cite{Bender:2002as,Bender:1996bb,Hellstern:1997nv}.
Nevertheless, correlations will persist in the diquark channels and
diquark masses found in ladder truncation can serve as a guide for
parametrizing those correlations.

\section{\label{sec:numcalc}
Numerical calculations}
\subsection{\label{sec:model}
Model interaction}
For our numerical calculations we employ a model for the effective
coupling ${\cal G}(k^2)$ that has been developed through an efficient
description of the masses and decay constants of the light
pseudoscalar and vector mesons~\cite{Maris:1999nt}.  This effective
$\bar q q$ interaction is constrained by perturbative QCD in the
ultraviolet and has a phenomenological infrared behavior.  The 
form is
\begin{eqnarray}
\label{gvk2}
\frac{{\cal G}(k^2)}{k^2} &=&
        \frac{4\pi^2\, D \,k^2}{\omega^6} \, {\rm e}^{-k^2/\omega^2}
        + \frac{ 4\pi^2\, \gamma_m \; {\cal F}(k^2)}
        {\textstyle{\frac{1}{2}} \ln\left[\tau + 
        \left(1 + k^2/\Lambda_{\rm QCD}^2\right)^2\right]} \;,
\end{eqnarray}
with \mbox{$\gamma_m=12/(33-2N_f)$} and \mbox{${\cal F}(s)=(1 -
\exp\frac{-s}{4 m_t^2})/s$}.  The ultraviolet behavior is chosen to be
that of the QCD running coupling $\alpha(k^2)$; the ladder-rainbow
truncation then generates the correct perturbative QCD structure of
the DSE-BSE system of equations.  The first term implements the strong
infrared support in the region \mbox{$0 < k^2 < 1\,{\rm GeV}^2$}
phenomenologically required~\cite{Hawes:1998cw} to produce a realistic
value for the chiral condensate.  We use \mbox{$m_t=0.5\,{\rm GeV}$},
\mbox{$\tau={\rm e}^2-1$}, \mbox{$N_f=4$}, \mbox{$\Lambda_{\rm QCD} =
0.234\,{\rm GeV}$}, and a renormalization scale \mbox{$\mu=19\,{\rm
GeV}$} which is well into the perturbative
domain~\cite{Maris:1997tm,Maris:1999nt}.  The remaining parameters,
\mbox{$\omega = 0.4\,{\rm GeV}$} and \mbox{$D=0.93\,{\rm GeV}^2$}
along with the quark masses, are fitted to give a good description of
the chiral condensate, $m_{\pi/K}$ and $f_{\pi}$.

Within this model, the quark propagator reduces to the one-loop
perturbative QCD propagator in the ultraviolet region.  In particular,
the dynamical mass function $M(p^2)=B(p^2)/A(p^2)$ behaves like
\begin{eqnarray}
 M(p^2) & \simeq & \frac{\hat{m}_q}{\left(
        \frac{1}{2}\ln\left[\frac{p^2}{\Lambda_{\rm QCD}^2}
                \right]\right)^{\gamma_m}} \;,
\end{eqnarray}
where $\hat{m}_q$ is the renormalization-point-independent explicit
chiral-symmetry-breaking mass.  In the chiral limit the
behavior is qualitatively different
\begin{eqnarray}
 M_{\hbox{\scriptsize{chiral}}}(p^2) & \simeq & \frac{2\pi^2\gamma_m}{3}\,
        \frac{-\,\langle \bar q q \rangle^0}{p^2 \left(
        \frac{1}{2}\ln\left[\frac{p^2}{\Lambda_{\rm QCD}^2}
                                        \right] \right)^{1-\gamma_m}}\,,
\label{eq:Mchiral}
\end{eqnarray}
with $\langle \bar q q \rangle^0$ the
renormalization-point-independent chiral
condensate~\cite{Maris:1997tm}.  Its relation to the
renormalization-dependent condensate
\begin{eqnarray}
  \langle \bar q q \rangle^\mu = - 3\,Z_4\int^\Lambda_q
        {\rm Tr}[S_{\hbox{\scriptsize{chiral}}}(p)] \,,
\end{eqnarray}
is at one-loop level
\begin{eqnarray}
\langle \bar q q \rangle^\mu &=& 
	(\ln{\mu/\Lambda_{\rm QCD}})^{\gamma_m} \, 
	\langle \bar q q \rangle^0 \,.
\end{eqnarray}
As demonstrated clearly in Ref.~\cite{Maris:1998hd}, this behavior of
the mass function in the chiral limit is a keystone of the microscopic
realization of Goldstone's theorem in QCD: the quark condensate (and
hence the pion mass) and other observables are materially dependent on
the ultraviolet properties of the interaction.

In the infrared region both the wave function renormalization $Z(p^2)
= 1/A(p^2)$ and the dynamical mass function deviate significantly from
the perturbative behavior, due to chiral symmetry breaking.  Recent
comparisons~\cite{Maris:2000zf,Tandy:2001qk} of results from this
rainbow DSE model to lattice QCD simulations~\cite{Skullerud:2001aw}
provide semiquantitative confirmation of the behavior generated by the
present DSE model: a significant enhancement of $M(p^2)$ and a
material enhancement of $A(p^2)$ below $1\;{\rm GeV}^2$.

The vector meson masses and electroweak decay constants produced by
this model are in good agreement with experiments~\cite{Maris:1999nt}.
The model's prediction for the pion charge form factor $F_\pi(Q^2)$
is confirmed by the most recent Jlab data~\cite{Volmer:2000ek}.
The kaon charge radii and electromagnetic form factors are also
described well~\cite{Maris:emf}, as is the weak $K_{l3}$
decay~\cite{Ji:2001pj}.  The strong decays of the vector mesons into a
pair of pseudoscalar mesons also agree reasonably well with
experiments~\cite{Maris:2001am,JMT01prep}.  The performance of this
model for deep inelastic scattering phenomena can be gauged from that
of a simplified version that has recently produced reasonable results
for the pion valence quark distribution amplitude~\cite{Hecht:2000xa}.

\subsection{Ground state mass spectrum of $q\bar{q}$ and $qq$ states}
With this model we have calculated the masses of the lightest two-body
bound states in the $u,d,s$ quark sector: pseudoscalar, scalar, and
vector mesons, and scalar, pseudoscalar, and axialvector diquarks.
As far as the flavor-singlet mesons are concerned, with the
rainbow-ladder kernel only flavor-singlet states with ``ideal mixing''
are possible.  This feature indicates a deficiency of the
rainbow-ladder truncation for studying flavor-singlet pseudoscalar and
scalar mesons (the flavor-singlet vector mesons are very close to
ideal mixing).

Regarding the flavor structure of the diquarks, the Pauli principle
prohibits the existence of flavor-singlet color-antitriplet scalar and
pseudoscalar diquarks; the only possible $ss$ diquarks are axialvector
diquarks~\cite{tuebingen,Alkofer:mv}.  Nevertheless, in model studies
such as this, one can consider the up/down quark masses to be equal to
the strange quark mass, and calculate the corresponding meson and
diquark masses.  This indicates the dependence of the diquark masses
on the current quark masses and allows for a comparison with other
calculations.  However, one should keep in mind that there are no
scalar nor pseudoscalar $ss$ diquarks in 3-flavor QCD with realistic
quark masses.

\begin{figure}[hbt]
\begin{center}
\epsfig{file=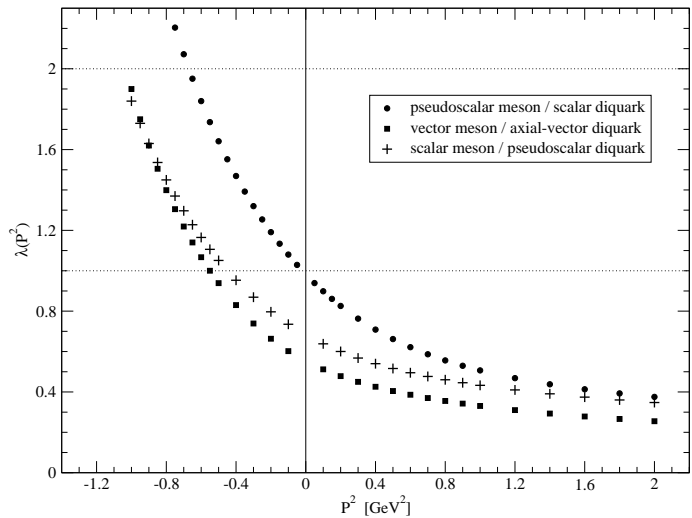,width=100mm}
\end{center}
\caption[]{The eigenvalue $\lambda(P^2)$ as function of $P^2$ in the
up/down sector: $\lambda=1$ signals a meson bound state, whereas
$\lambda=2$ corresponds to a diquark bound state. \label{fig:lambda}}
\end{figure}

The bound state BSE has solutions at discrete values of $P^2 = -m^2$,
where $m$ is the mass of the bound state.  In order to determine the
bound state masses, we introduce a parameter $\lambda(P^2)$, and turn
the BSE into an eigenvalue problem for $\lambda(P^2)$
\begin{eqnarray}
 {\textstyle\frac{4}{3}}\int_q^\Lambda\!\! {\cal G}(k^2) \,
	D_{\mu\nu}^{\rm free}(k) \, \gamma_\mu \, S(q_+) \, 
	\Gamma(q_+,q_-) \, S(q_-) \, \gamma_\mu  
	& = & \lambda(P^2) \, \Gamma(p_+,p_-) \,.
\label{eq:ladderBSE}
\end{eqnarray}
This equation has solutions for all $P^2$, but in general these
solutions do not correspond to bound state solutions of the original
BSE, Eqs.~(\ref{eq:ladderBSEm}) and ~(\ref{eq:ladderBSEd}).  Meson
bound states with mass $m$ are described by solutions of
Eq.~(\ref{eq:ladderBSE}) $\Gamma(p_+,p_-) = \Gamma_M(p_+,p_-)$ with
$\lambda(P^2=-m^2) = 1$, whereas diquark bound states are described
by solutions $\Gamma(p_+,p_-) = \Gamma_D(p_+,-p_-) \,C $ with
$\lambda(-m^2) = 2$ (remember that the color structure in the diquark
BSE leads to a factor $\frac{2}{3}$ instead of $\frac{4}{3}$ , see
Eq.~(\ref{eq:diqcol})).  In Fig.~\ref{fig:lambda} we display the
eigenvalues as function of $P^2$ in the up/down sector.  From this
figure we can simply read of the values of $P^2$ that give us the mass
of a bound state.

Since we started from a Euclidean formulation, we have to make an
analytic continuation to negative values of $P^2$ in order to
determine the bound state solutions.  For small masses, this is no
problem; however, for larger masses this becomes numerically more and
more cumbersome.  Moreover, the quark propagator in rainbow truncation
often has singularities at complex values of $p^2$~\cite{sing}, which
could be interpreted as a signal for confinement~\cite{conf}.  Once
the integration domain sampled in the BSE includes these
singularities, one has to specify how to deal with these
singularities.  Alternatively, one can extrapolate the eigenvalues
$\lambda(P^2)$ from the region where the analytic continuation is
unambiguous to $\lambda=1$ for mesons, and to $\lambda=2$ for
diquarks, whenever necessary.  We followed the latter procedure, using
several different extrapolating functions to reduce the numerical
uncertainty.  For the bound states with the largest mass, we estimate
the error due to this extrapolation to be a few percent.

\begin{table}[hbt]
\footnotesize
\begin{tabular}{lccc@{\hspace{7mm}}ccc@{\hspace{7mm}}ccc}
\firsthline
	& \multicolumn{3}{l}{$m_{q} = m_{q'} = m_{u,d}$}
	& \multicolumn{3}{l}{$m_{q} = m_{u,d}$, $m_{q'} = m_s$}
	& \multicolumn{3}{l}{$m_{q} = m_{q'} = m_s$} \\ \midhline
meson	& $\pi$	& $\rho,\omega$	& $\sigma$ 	
	& $K$ 	& $K^\star$	& $\kappa$ 	
	& $0^-$ & $\phi$	& $0^+$	\\ \midhline
calc. 	& 0.138	& 0.741	& 0.671
	& 0.496	& 0.937	& 0.893		
	& 0.696	& 1.07	& 1.08		\\
dom.    & 0.121 & 0.875	& 0.759
	& 0.425	& 1.08	& 1.03		
	& 0.588	& 1.24	& 1.30		\\
sep.    & 0.139 & 0.736	& 0.715		
	& 0.494	& 0.854	&  --		
	& 	& 0.950	&		\\
lat.	& 0  	& 0.579	&		
	&	&	&		
	& 0.910	& 1.025	&		\\ \midhline
diquark	& $0^+$	& $1^+$	& $0^-$	
	& $0^+$	& $1^+$	& $0^-$ 
	& $0^+$	& $1^+$	& $0^-$		\\ \midhline
calc.	& 0.82	& 1.02	& 1.03		
	& 1.10	& 1.30(6) & 1.31(4)		
	& 1.27	& 1.44(4) & 1.50(4)	\\  
dom.    & 0.74  & 1.06	& 1.14(2)		
	& 0.94	& 1.34(4) & 1.45(4)
	& 1.12	& 1.51(4) & 1.72(6)	\\
sep.    & 0.74  & 0.95	& 1.50		
	& 0.88	& 1.05	& --		
	& 	& 1.13	&		\\
lat.	& 0.62	& 0.73	&		
	&	&	&		
	& 1.19	& 1.21	&		\\ \lasthline
\end{tabular}
\caption[]{\label{tab:mass}
\footnotesize
Overview of our results for the meson and diquark masses for different
current quark masses, all in GeV, compared with the results from
a separable model~\protect{\cite{Burden:1996nh}} and with quenched
lattice QCD~\protect{\cite{Hess:1998sd}}.  Note that the Pauli
principle prohibits the existence of flavor-singlet color-antitriplet
scalar and pseudoscalar diquarks; the results for the \protect{
$m_q = m_{q'} = m_s$} scalar and pseudoscalar diquark masses are only 
included to indicate the dependece on the current quark mass and for
comparison with lattice data.  Our numerical errors are of the order
of 1\%, with the exception of the heavier-mass states, which have an
extrapolation error as indicated.}
\end{table}
Our results are displayed in Table~\ref{tab:mass}, and compared with
results from rainbow-ladder truncation of the DSEs using a separable
Ansatz for the interaction~\cite{Burden:1996nh} and with lattice
data~\cite{Hess:1998sd} where available.  Note that our calculation is
done at realistic quark masses, whereas the lattice results are
extrapolated to the chiral limit.  For the strange quark we have taken
the lattice data with $m_V = m_\phi$.  Compared to the lattice data,
we find a larger splitting between the scalar and the axialvector
diquark; also in the $s\bar{s}$ sector we find a significantly larger
splitting between the pseudoscalar and vector meson.  The lattice
simulations do not report on scalar meson or pseudoscalar diquark
masses.

Our results are qualitatively the same as the results of the separable
model of Ref.~\cite{Burden:1996nh}, with the exception of the scalar
mesons and pseudoscalar diquarks.  We find the scalar and vector meson
masses to be within 10\% of each other, not only in the $u/d$ sector,
but also in the $us$ and $ss$ sectors.  Similarly, we also find the
pseudoscalar diquark and axialvector diquark masses to be within about
10\% of each other, whereas in the separable model the pseudoscalar
diquark is much heavier than the axialvector diquark, or not bound at
all.  Also notice that the scalar meson and pseudoscalar diquark
masses increase more rapidly with the quark mass than the vector meson
and scalar diquark masses: e.g. the $m_\sigma < m_\rho$ but
$m_{0^+}^{s\bar{s}} > m_\phi$.

In Table~\ref{tab:mass} we also give the masses obtained using only
the dominant covariant for the different bound states; e.g., retaining
only $E$ in Eq.~(\ref{genpion}).  This approximation changes the mass
by about 20\% both for the mesons and for the diquarks.  Using the
dominant covariants only tends to increase the difference between
pseudoscalar meson masses and vector and scalar masses; we see a
similar increase in the difference in the diquark channels.

The meson masses seem to be rather independent of the details of the
effective interaction, as long as the interaction generates the
observed amount of chiral symmetry breaking, as can be seen from
Table~\ref{tab:paramdep}.  The parameters $\omega$ and $D$ were fitted
in Ref.~\cite{Maris:1999nt} to reproduce $f_\pi$ and the chiral
condensate.  Within this parameter range, the vector meson masses are
almost independent of the parameters, and the scalar meson masses
change by only 10\%.  The diquark masses are more sensitive to details
of the effective interaction.  Nevertheless, the lightest diquark is
the scalar diquark, independent of the parameters, with the
axialvector and pseudoscalar diquarks being about 150 to 250 MeV
heavier.
\begin{table}[hbt]
\footnotesize
\begin{tabular}{lccc@{\hspace{12mm}}ccc}
\firsthline
&
\multicolumn{3}{c}{mesons} &
\multicolumn{3}{c}{diquarks}\\
	& $\pi$	& $\rho,\omega$	& $\sigma$ 	
	& $0^+$	& $1^+$	& $0^-$		\\ \midhline
$\omega=0.3\;{\rm GeV}$, $D=1.25\;{\rm GeV}^2$ 
	& 0.139	& 0.746	& 0.669
	& 0.98	& 1.1	& 1.2		\\
$\omega=0.4\;{\rm GeV}$, $D=0.93\;{\rm GeV}^2$ 
	& 0.138	& 0.741	& 0.671
	& 0.82	& 1.02	& 1.03		\\
$\omega=0.5\;{\rm GeV}$, $D=0.79\;{\rm GeV}^2$ 
	& 0.138	& 0.742	& 0.596
	& 0.688	& 0.89	& 0.86		\\ \lasthline
\end{tabular}
\caption[]{\label{tab:paramdep}
\footnotesize
Parameter dependence of the masses (in GeV) in the up/down sector.}
\end{table}
%

\subsection{Bethe--Salpeter amplitudes}
\begin{figure}[hbt]
\begin{center}
\epsfig{file=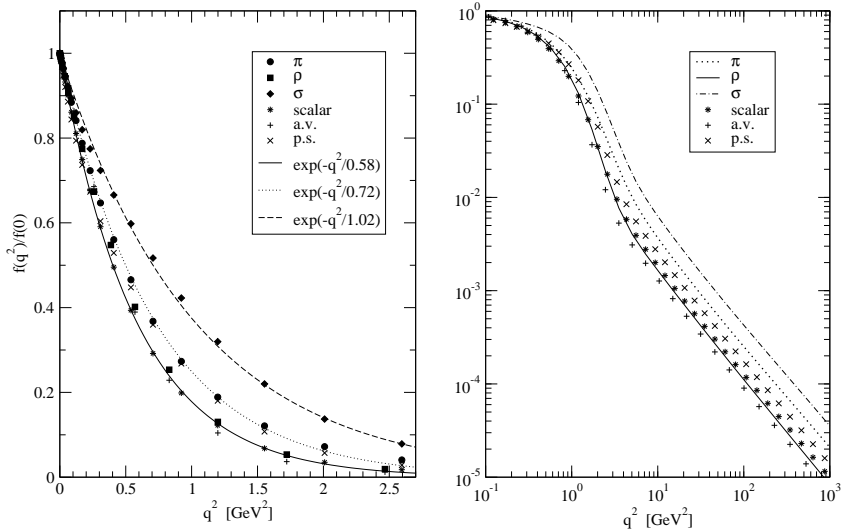,width=120mm}
\end{center}
\caption[]{Leading Chebyshev moments of canonical amplitudes in the
up/down sector.\label{fig:BSAuu}}
\end{figure}

The corresponding BSAs are similar for the different bound states.  In
Fig.~\ref{fig:BSAuu} we have displayed the leading Chebyshev moments
\begin{eqnarray}
  f(q^2)  &=& \int_0^\pi \sin^2\theta \; d\theta  \; 
		f(q^2,\,q P\cos\theta;\eta=\textstyle{\frac{1}{2}}) 
\end{eqnarray}
of the amplitudes associated with the canonical Dirac structure for
the various mesons and diquarks in the up/down sector, all normalized
to $f(0) = 1$: $E$ for pseudoscalar and scalar bound states, see
Eq.~(\ref{genpion}), and $V_1$ for vector and axialvector bound
states, see Eq.~(\ref{vecBSAdom}).

In the infrared region, $0 < q^2 < 2\;{\rm GeV}^2$, the canonical
covariants can reasonably well be approximated by exponentials
$f(q^2)=\exp(-q^2/\omega^2)$ with $0.74\;{\rm GeV} < \omega <
1.01\;{\rm GeV}$, depending on the meson or diquark, see
Table~\ref{tab:bsas}.
\begin{table}[hbt]
\footnotesize
\begin{tabular}{lccc@{\hspace{15mm}}ccc}
\firsthline
&
\multicolumn{3}{c}{mesons} &
\multicolumn{3}{c}{diquarks}\\
state	& $\pi$	& $\rho$& $\sigma$	& $0^+$	& $1^+$	& $0^-$	\\
$\omega$ [GeV] & 0.84  & 0.79  & 1.01	& 0.76  & 0.74  & 0.82 	\\
							\\ \lasthline
\end{tabular}
\caption[]{\label{tab:bsas}
\footnotesize
Comparison of the BSAs for the up/down states, fitted to 
$\exp(-q^2/\omega^2)$ in the infrared region.}
\end{table}
At large $q^2$, they all fall off like $c/q^2$ up to calculable
logarithmic corrections~\cite{Maris:1997tm,Maris:1999nt}.  It is thus
apparent that in all three channels, the diquark BSAs are narrower in
momentum space than the corresponding meson BSAs, and thus wider in
coordinate space.  This is what one would expect, since the diquarks
are less bound.  It is intriguing that the scalar meson BSA is broader
in momentum space than the pion BSA, and thus narrower in coordinate
space.  One expects the {\em physical} $\sigma$ state to include
significant pion cloud contributions and to thus be significantly
broader in coordinate space than the pion.  In ladder BSE
approximation, such physical contributions are not accommodated; also
other terms beyond rainbow-ladder truncation are known to be important
in the scalar channel.  Clearly, additional study beyond ladder
truncation is required in the scalar channel.

\section{Concluding remarks}
We have calculated the meson and color-antitriplet diquark masses
using DSEs in rainbow-ladder truncation.  The model parameters have
previously been fixed by the calculation of $f_\pi$, $m_\pi$, $m_K$,
and the chiral condensate, and it leads to a good description of the
light pseudoscalar and vector mesons and their interactions.  Using
this model, we find that the lightest diquark bound states in a color
antitriplet configuration is a scalar diquark, with a mass of about
800 MeV in the case of up and down quarks.  The axialvector and
pseudoscalar diquarks are about 150 to 250 MeV heavier than the scalar
diquark.  In the meson channel, we find a scalar meson with a mass
slightly below the vector meson for up/down quarks, but slightly above
the vector meson mass for strange quarks.  These results are
consistent with the notion of a (broad) $\sigma$-meson with a mass
around 600 MeV, and with the mass splitting between the scalar and
vector mesons in heavy quarkonia.  The corresponding BSAs are all
similar in shape, and can in the infrared region be described by a
gaussian.  For large relative momenta, they fall off like $1/p^2$,
with calculable logarithmic corrections which is characteristic of
QCD.

The ladder truncation is particularly suitable for the pseudoscalar
mesons and for the vector mesons, because of cancellations in the
higher-order contributions to the kernel.  However, diquarks, and also
the scalar meson, are expected to receive significant corrections from
higher order terms in the kernel.  This can significantly change the
bound state mass and corresponding BSA; in particular the diquarks
will no longer be a true bound state.  Thus one should only consider
these masses as an indication for the relevant mass scales of the
diquark correlations.  As such they should be useful in e.g. studies
of baryons as a quark-diquark system.

Typically, studies of baryons as a quark-diquark bound state include
both scalar and axialvector diquarks in a color antitriplet, but no
pseudoscalar diquarks.  The present study indicates that the mass
scale of pseudoscalar diquarks is similar to that of axialvector
diquarks.  It may therefore be interesting to explore the influence of
pseudoscalar correlations on quark-diquark models for baryons such as
those of Refs.~\cite{rtcahill,bentz,tuebingen,argonne}.

\section*{Acknowledgements}
I would like to thank Craig Roberts and Peter Tandy for useful
discussions.  This work is supported by DOE under grants
No. DE-FG02-96ER40947 and DE-FG02-97ER41048 and benefitted from 
the computer resources at NERSC.


\end{document}